\documentclass[10pt,twocolumn,letterpaper]{article}

 \usepackage{cvpr}              % To produce the CAMERA-READY version
\usepackage[accsupp]{axessibility}  % Improves PDF readability for those with disabilities.
\usepackage{graphicx}
\usepackage{amsmath}
\usepackage{amssymb}
\usepackage{booktabs}
\usepackage{bbold}
\usepackage{enumitem}
\usepackage{tabularx}

\usepackage[pagebackref,breaklinks,colorlinks]{hyperref}
\DeclareUnicodeCharacter{2212}{-}

\usepackage[capitalize]{cleveref}
\crefname{section}{Sec.}{Secs.}
\Crefname{section}{Section}{Sections}
\Crefname{table}{Table}{Tables}
\crefname{table}{Tab.}{Tabs.}

\def\cvprPaperID{35} % *** Enter the CVPR Paper ID here
\def\confName{CVPR}
\def\confYear{2023}

\begin{document}

% \title{Contrastive regularized data Fusion for robust guided Thermal Super Resolution}
\title{CoReFusion: Contrastive Regularized Fusion for Guided Thermal Super-Resolution}

\author{Aditya Kasliwal, Pratinav Seth, Sriya Rallabandi\thanks{Authors have contributed equally to this work} , Sanchit Singhal\footnotemark[1]\\
Manipal Institute of Technology\\ Manipal Academy of Higher Education, Manipal, India\\
 \texttt{\{kasliwaladitya17,seth.pratinav,sriyarallabandi,sanchitsinghal\}@gmail.com}}
\maketitle
%%%%%%%%% ABSTRACT
\begin{abstract}

Thermal imaging has numerous advantages over regular visible-range imaging since it performs well in low-light circumstances. Super-Resolution approaches can broaden their usefulness by replicating accurate high-resolution thermal pictures using measurements from low-cost, low-resolution thermal sensors. Because of the spectral range mismatch between the images, Guided Super-Resolution of thermal images utilizing visible range images is difficult. However, In case of failure to capture Visible Range Images can prevent the operations of applications in critical areas. 
We present a novel data fusion framework and regularization technique for Guided Super Resolution of Thermal images. The proposed architecture is computationally in-expensive and lightweight with the ability to maintain performance despite missing one of the modalities, i.e., high-resolution RGB image or the lower-resolution thermal image, and is designed to be robust in the presence of missing data. The proposed method presents a promising solution to the frequently occurring problem of missing modalities in a real-world scenario. Code is available at \url{https://github.com/Kasliwal17/CoReFusion}.

\end{abstract}

%%%%%%%%% BODY TEXT
\section{Introduction}
\label{sec:intro}
RGB Images have been important for years as a primary means of capturing and displaying visual information, but their inability to capture information relevant to infrared radiation has made thermal images preferable to capture objects beyond the human spectrum. Thermal images are captured by thermal cameras, which are passive sensors that detect infrared light at temperatures greater than absolute zero. These images can detect temperature variations that are not visible to the human eye, work in low or even no-light conditions, don't need any external illumination, and are not affected by environmental conditions such as snow, fog, or haze as a result of which they are useful in detecting sources of heat or other applications in such conditions. These unique advantages make them particularly useful in construction, agriculture, medicine, law enforcement, and military operations \cite{Gade2013ThermalCA}.\\
Despite their popularity, the costs of using a quality thermal camera that gives acceptable resolution, good accuracy, and less noise interference have improved image resolution from a hardware level. As a result, the software-based approach of Single Image Super-Resolution (SISR)\cite{dong2015image} has emerged as a viable option to enhance image resolution in thermal imaging. SISR\cite{yang2014single} is primarily a class of image processing techniques used to enhance the resolution of digital images. It is particularly useful in situations where the original image is low-resolution or pixelated. These algorithms operate by learning a mapping function between low-resolution and high-resolution images, typically using deep neural networks. These techniques when used for thermal imagery, called Thermal Image Super Resolution(TISR)\cite{inproceedings}, is particularly useful in situations where the original thermal image has a low spatial resolution or is pixelated. With the growing use of low-cost thermal cameras, there is a growing interest in this technique that can enhance the visual quality of thermal images captured by these devices, thereby reducing the dependence on high-quality hardware and creating a more equitable option.\\
However, with low-quality thermal images, Guided Thermal Image Super Resolution \cite{8713356} enhances the image resolution by utilizing low-resolution thermal images and high-resolution RGB images as inputs to generate a high-quality thermal image. This method leverages the complementary information between low-resolution thermal and high-resolution RGB images. RGB images contain more detailed information about edges and structures, while thermal images provide valuable information about temperature distribution. Guided TISR algorithms\cite{9924742} can use both images to generate high-quality thermal images with improved spatial resolution, sharper edges, and more detailed structures.
\\
These advantages of Guided TISR have led to a rise in interest in this domain among researchers belonging to academia, industry, and government agencies. As a result, the 19th IEEE Workshop on Perception Beyond the Visible Spectrum introduced a challenge track this year to generate x8 super-resolution thermal image by using a high-resolution visible image as guidance for a low-resolution thermal image, using a newly generated dataset acquired with cross-spectral sensors such as Balser and TAU2 camera.\\
In this paper, we put forth our solution to the challenge, a UNet-based architecture\cite{DBLP:journals/corr/RonnebergerFB15} with two encoders, similar to the approach proposed in \cite{https://doi.org/10.48550/arxiv.2009.10608}. It is a lightweight and novel data fusion model that is robust and computationally efficient and a promising solution for dealing with missing modalities.
%-------------------------------------------------------------------------
\section{Related Works}
\subsection{Visible Image Super Resolution}
Visible image super-resolution (VISR) has been a popular research topic for over a decade. Early methods include model-based methods like neighbour embedding regression\cite{6751349} and Random Forest\cite{7299003}. In the past few years, CNN-based models have become popular due to their high performance in computer vision tasks.SRCNN\cite{dong2015image} is the first CNN-based SISR model extracts features from the LR images. It learns the mapping between LR and HR features and reconstructs their HR images. Inspired by this, some advanced models were made in this model, like FSRCNN\cite{dong2016accelerating}, EDSR\cite{lim2017enhanced}, and VDSR\cite{7780551}. A few of them also included unsupervised learning, such as DRN\cite{guo2020closed} and DASR\cite{wang2021unsupervised}. Some recent work combined with vision transformer (ViT), such as SwinIR\cite{liang2021swinir}.
\subsection{Thermal Image Super Resolution}
The achievement of deep learning in VISR has encouraged further research in Thermal image super-resolution (TISR).\cite{inproceedings} proposed a GAN-based novel architecture called CycleGAN along with a large thermal image dataset.\cite{9924742} proposes a method to increase the resolution of thermal images using edge features of corresponding high-resolution visible images.\cite{Prajapati2021ChannelSC} proposed a Channel Splitting-based Convolutional Neural Network (ChasNet) for thermal image SR eliminating the redundant features in the network.
\subsection{Guided Thermal Image Super Resolution}
Guided TISR is the method for super-resolution of thermal images guided by the information integrated from the RGB image counterpart.
\cite{8713356} proposed a GAN-based model which enhances the thermal image resolution using the counter RGB images. The outcomes demonstrate that RGB-guided thermal super-resolution models enhance resolution compared to traditional single thermal super-resolution techniques.\cite{gupta2020pyramidal}
proposed a GSR based on pyramidal edge maps extracted from the visible image. The proposed network has two sub-networks. The first sub-network super-resolves the low-resolution thermal image, while the second obtains edge maps from the visible image at a growing perceptual scale and integrates them into the super-resolution sub-network with the help of attention-based fusion.
\begin{figure*}[ht]
  \centering
  \includegraphics[width=\linewidth]{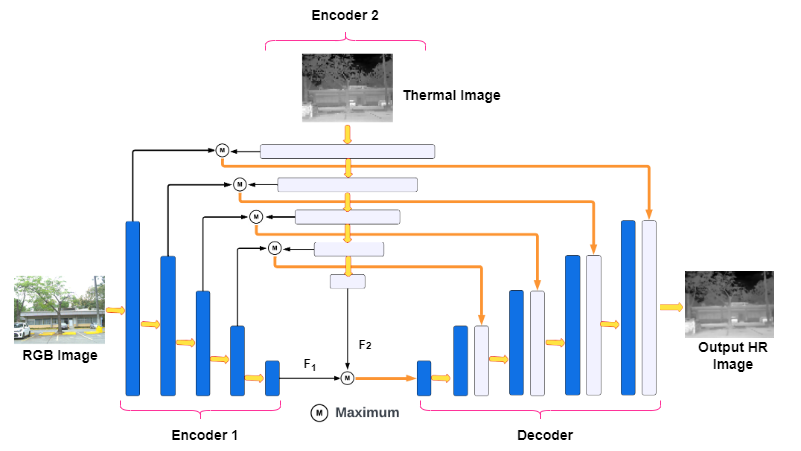}
  \caption{Proposed CoReFusion model architecture}
  \label{figure 1}
\end{figure*}
\subsection{Contrastive Learning}
Contrastive learning is a self-supervised learning technique that improves the performance of models by using the principle of contrasting samples against each other to learn attributes that are common between data classes and attributes that set apart a data class from another.\cite{chen2020simple} proposed SimCLR: a simple framework for contrastive learning of visual representations which outperform previous methods for self-supervised and semi-supervised learning on ImageNet.\cite{wu2021practical} proposed a practical contrastive learning framework for SISR that involves the generation of many informative positive and hard negative samples in frequency space. Instead of utilizing an additional pre-trained network, they also design a simple but effective embedding network inherited from the discriminator network and can be iteratively optimized with the primary SR network making it task-generalizable. \cite{9856966}, proposed a Camera Internal Parameters Perception Network (CIPPSRNet) for LR thermal image enhancement. They also introduced contrastive learning to optimize the pretrained Camera Internal Parameters Representation Network and the feature encoders to improve the network’s adaptability to misalignment data with insufficient pixel matching and robustness. Little work has been done in super-resolution using contrastive learning, leaving this field unexplored.
\subsection{Missing Modality in Multi-Modal Learning}
Multi-modal learning has risen over the past decade, especially in Computer Vision. This involves the development of models capable of processing and analyzing multi-modal information uniformly. \cite{Bayoudh2021ASO} conveyed key concepts and algorithms in Deep multi-modal learning for computer vision by exploring how to generate deep models that consider the integration and combination of heterogeneous visual cues across sensory modalities with a discussion about various limitations of multi-modal learning, like real-time processing, missing modality, and data conflict. \cite{ma2021smil} did a study on multi-modal learning with missing modality in training, testing, or both and its efficiency. They also propose a new SMIL method that leverages Bayesian meta-learning to achieve both objectives uniformly.

\cite{woo2022towards} tried to explore various reliable methods for multi-modal action recognition, focusing on solutions where some modalities are missing at an inference. They explored various techniques for fusion that can improve the ability to handle situations where some modalities are missing and discovered that fusion methods based on transformers exhibit stronger resilience towards missing modalities than summation or concatenation methods. They proposed a modular network, ActionMAE \cite{woo2022towards}, which learns missing modality predictive coding by randomly dropping modality features and tries to reconstruct them with the remaining modality features.

\section{Proposed Method}
In this paper, we introduce a novel methodology as a solution for the PBVS 2023 TISR track-2 challenge, which involves generating high-resolution thermal images by utilizing a low-resolution thermal image with the guidance of a high-resolution visible RGB image. Our proposed approach, illustrated in figures \ref{figure 1} and \ref{figure 2}, comprises two primary components:
\begin{itemize}[itemsep=0pt]
    \item An U-Net based Model architecture incorporating two ResNet-34 encoders
    \item Two Contrastive modules that receive the final features $F_1$ and $F_2$ from both encoders
\end{itemize}
We propose a modified U-Net\cite{DBLP:journals/corr/RonnebergerFB15} based architecture incorporating two ResNet-34\cite{DBLP:journals/corr/HeZRS15} encoders and a fusion method for combing their features. Similar to the U-Net architecture, the encoders down-sample the input image and extract features at different levels of abstraction. However, the two encoders work independently to extract features. The first encoder is designed to receive the high-resolution RGB image, while the second encoder receives the low-resolution thermal image that has been up-sampled using bi-linear interpolation to a size comparable to that of the high-resolution RGB image.
\subsection{Feature Transfer from Encoders to Decoder}
%draft 1
%  After the features are extracted from both encoders, a new feature map is calculated by taking the elementwise maximum of the similar level feature maps from both encoders. This process ensures that the most relevant features are selected from each encoder and combined to form a new feature map that captures both the thermal and visible information.
% The new feature map is then transferred to the decoder using skip connections. The decoder, which is also composed of several decoding blocks, upsamples the feature maps and combines them with the corresponding feature maps from the encoders using the skip connections. However, in this modified Unet model, the skip connections are used to transfer the new feature map that was calculated from the two encoders, instead of the individual feature maps from each encoder.
% By combining the features from both the RGB and thermal images at multiple levels of abstraction, the modified Unet model is able to produce high-quality outputs.
% draft 2
Features after being extracted from both encoders, the most relevant features are selected from each encoder and combined to form a new feature map that captures both the thermal and visible information. This is achieved by taking the element-wise maximum of the feature maps at each similar level of abstraction from both encoders. The resulting new feature map is then transferred to the decoder using skip connections.
\\
The decoder, composed of several decoding blocks, up-samples the feature maps and combines them with the corresponding feature maps from the encoders using skip connections. In our proposed model, the skip connections are used to transfer the new feature map calculated from the two encoders rather than the individual feature maps from each encoder. By combining the features from both the RGB and thermal images at multiple levels of abstraction, our model is able to produce high-quality outputs.

\subsection{Contrastive Regularization}
\begin{figure}[ht]
  \centering
  \includegraphics[width=\linewidth]{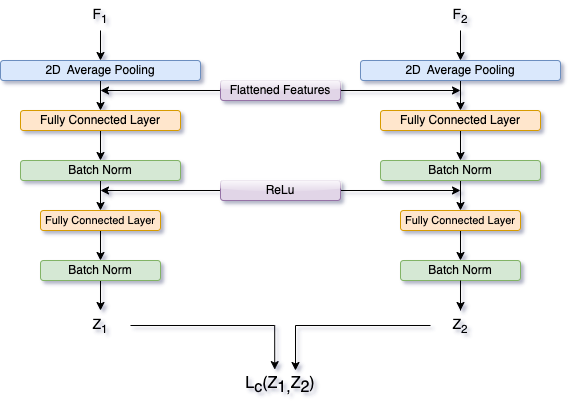}
  \caption{Contrastive Modules}
  \label{figure 2}
\end{figure}

% The final features F1 and F2 that are extracted from the encoders are transferred to the two contrastive modules respectively. Each contrastive module as shown in fig 2 has a series of layers: an average pooling layer, a fully connected layer ,batch normalization and then again an fully connected layer followed by batch normalization. The output Z1 and Z2 from these contrastive heads is used to calculate the contrastive loss, which is then added to our final loss function after being multiplied by a hyperparameter beta.
The features $F_1$ and $F_2$ extracted from the encoders as depicted in Figure \ref{figure 1} are transferred to two separate contrastive modules. Figure \ref{figure 2} shows that each contrastive module includes layers such as average pooling, fully connected, and batch normalization, followed by another fully connected layer and batch normalization. 
The resulting outputs, $Z_1$ and $Z_2$, from the contrastive modules, are utilized to compute the contrastive loss which is scaled using the hyper-parameter $\beta$ and is then added to our final loss.

%which is subsequently added to our final loss function after being multiplied by hyper-parameter $\beta$.After computing the contrastive loss using Z1 and Z2, we adjust the value by scaling it with the hyper-parameter $\beta$. The resulting scaled value is then added to our final loss function.After computing the contrastive loss using $Z_1$ and $Z_2$, We scale the loss using the hyper-parameter $\beta$, which is then added to our final loss.
\subsection{Missing Modality}
In case of missing modalities, such as the absence of guiding High-resolution RGB image in the input, the model can still perform well due to the fact that the two encoders work independently. In such scenarios, the model will utilize only one encoder, and the multilevel feature maps from this encoder will be directly passed to the decoder through the skip connections. This mechanism guarantees the model can still obtain relevant features from the available input and generate accurate outputs.
% \subsection{Mathematics}

% Please number all of your sections and displayed equations as in these examples:
% \begin{equation}
%   E = m\cdot c^2
%   \label{eq:important}
% \end{equation}
% and
% \begin{equation}
%   v = a\cdot t.
%   \label{eq:also-important}
% \end{equation}
% It is important for readers to be able to refer to any particular equation.
% Just because you did not refer to it in the text does not mean some future reader might not need to refer to it.
% It is cumbersome to have to use circumlocutions like ``the equation second from the top of page 3 column 1''.
% (Note that the ruler will not be present in the final copy, so is not an alternative to equation numbers).
% All authors will benefit from reading Mermin's description of how to write mathematics:
% \url{http://www.pamitc.org/documents/mermin.pdf}.
\subsection{Loss Function}

\begin{table*}[h!]
\begin{tabularx}{\textwidth }{XXXXXX}
\hline
$\boldsymbol{\beta}$ & \textbf{Epoch Range} &\textbf{Train SSIM} & \textbf{Val. SSIM} & \textbf{Train PSNR} & \textbf{Val. PSNR} \\
\hline
0 & 375 $\pm{ 25}$ & 0.8705 & 0.8300 & 30.407 & 27.827\\
0.001 & 235 $\pm{10}$ & 0.8487 & 0.8308 & 29.335 & 27.691 \\
0.01 & 255 $\pm{10}$ & 0.8397 & 0.83 & 28.930 & 27.817 \\
0.1 & 360 $\pm{10}$ & 0.8403 & 0.8347 & 28.625 & 27.992 \\
1 & 455 $\pm{15}$ & 0.8300 & 0.8322 & 28.190 & 28.039 \\
\hline
\end{tabularx}
\caption{Train and Validation(Val.) Results for Various values of $\beta$ ( Hyperparameter for Contrastive Loss) with SSIM and PSNR metrics}
\label{tab:Table 1}
\end{table*}

\subsubsection{Mean Square Error(MSE) Loss}
MSE loss compares how far away the target image’s pixels are from the predicted/generated image’s pixels. Minimizing the loss tries to make the predicted/generated image as close as possible to the ground truth image in terms of the pixel values. Generally, Mean Absolute Error (MAE) is used in image enhancement as a loss function that minimizes MSE and encourages finding pixel averages of solutions for smoothing the images. And even when the loss is minimised, the generated images will have poor perceptual quality from the perspective of appealing to a human viewer. But through careful experimentation, we observed that MSE loss performs better than MAE loss when used with contrastive loss.
\begin{equation}
L_{MSE} = \frac{1}{n} \sum_{i=1}^{n}(y_i - \hat{y}_i)^2
\end{equation}

where $n$ is the number of observations, $y_i$ is the true value of the $i^{th}$ observation, and $\hat{y}_i$ is the predicted value of the $i^{th}$ observation.
\subsubsection{Peak signal-to-noise ratio (PSNR)}
PSNR describes the spatial reconstruction in the generated images and is defined by the ratio between the signal's maximum power and the power of the residual errors.
Specifically, it measures the amount of information loss between the original and reconstructed image by calculating the mean squared error (MSE) and then transforming it to a decibel (dB) scale using the formula:
\begin{equation}
PSNR = 10 \log_{10} \left( \frac{MAX^2}{MSE} \right)
\end{equation}

where $MAX$ is the maximum possible pixel value of the image, and $MSE$ is the mean squared error between the original and reconstructed image.
\begin{equation}
L_{PSNR} = \frac{1}{n} \sum_{i=1}^{n}(1 - (PSNR/40))
\end{equation}
PSNR is divided by 40 to bring it to a scale similar to SSIM.
\\
A higher PSNR value implies a greater quality of the spatial reconstruction in the final image. If the images are identical, PSNR is equal to infinity.
\subsubsection{Structural Similarity Index(SSIM)}
SSIM\cite{psnr} is a metric used to measure the similarity between two images. It takes into account three factors: luminance, contrast, and structure. SSIM is widely used \cite{huang2019single,nay2021single} as a loss function in image super-resolution tasks as the model using MSE output images during its training process viewed by human perception tends to be a lot blurrier and more undefined compared to SSIM.
\begin{equation}
\text{SSIM}(x,y) = \frac{(2\mu_x\mu_y + C_1)(2\sigma_{xy} + C_2)}{(\mu_x^2 + \mu_y^2 + C_1)(\sigma_x^2 + \sigma_y^2 + C_2)}
\end{equation}

where $x$ and $y$ are the input images being compared
$\mu_x$ and $\mu_y$ are the means of $x$ and $y$, respectively
$\sigma_x$ and $\sigma_y$ are the standard deviations of $x$ and $y$, respectively
$\sigma_{xy}$ is the cross-covariance of $x$ and $y$,   
$C_1$ and $C_2$ are constants used to stabilize the division with a weak denominator
\\
\begin{equation}
L_{SSIM} = \frac{1}{n} \sum_{i=1}^{n}(1 - SSIM)
\end{equation}
The SSIM loss function encourages the output image to have similar luminance, contrast, and structure as the ground truth image. By minimizing this loss, the model can learn to generate high-quality super-resolved images visually similar to the ground truth images.
 
\subsubsection{Contrastive Loss}
Contrastive loss is a loss function used in machine learning to train models for similarity learning. Similarity learning aims to learn a representation of data so that similar data points are closer together in the representation space while dissimilar data points are farther apart. It helps to reduce artifacts and noise in the output images, resulting in higher fidelity and more realistic images.
\cite{chen2020simple} used this function to train the model by comparing pairs of images and computing a similarity score between them.
\begin{equation}
l(v_i,v_j) = -\log\left(\frac{\exp(\text{sim}(v_i, v_j))}{\sum_{k \neq i} \exp(\text{sim}(v_i, v_k))}\right)
\end{equation}
where $v_i$ and $v_j$ are the representations of two augmented views of the same input image, $\text{sim}(·, ·)$ is a function calculating cosine similarity, and the denominator is the sum over all views $k$ in the batch except for $i$.
\\
Finally, the loss is computed over all the pairs in the batch of size N and takes an average.
\\
\begin{equation}
L_{C} = \frac{1}{n} \sum_{k=1}^{n}[l(2k-1,2k)+l(2k,2k-1)]
\end{equation}
\subsubsection{Final Loss}
% The final loss used in our paper after careful experimentation with different hyperparameters is given by
The final loss used in our paper, determined through meticulous testing of a combination of various hyper-parameters, is given by
\begin{equation}
L_{Final} = L_{MSE} + 0.1*L_{PSNR} + 0.1*L_{SSIM} + \beta*L_{C} 
\end{equation}
where $\beta$ represents a hyperparameter that regulates the extent of contrastive loss desired.

\section{Experiments}
We use a new dataset provided by the challenge organizers\cite{upsaclayCodaLabCompetition}, consisting of registered pairs of high-resolution visible and low-resolution thermal images of the same scene in 640x480 resolution, captured using cross-spectral sensors Balser and TAU2 camera. We used 190 pairs of such images provided to conduct our experiments, with 160 used for training and 30 for validation, and the remaining 10 for the testing phase by the organizers. We conducted all of our experiments on the validation dataset only due to the non-availability of the test dataset, as it was a part of the Thermal Image Super-Resolution challenge (Track 2).

During our experimentation, we analyzed various activation functions, such as Sigmoid, ReLU, and Tanh, for the output layer of the model. Our observations revealed that Sigmoid was ineffective as it significantly impacted extreme values. We observed that ReLU produced favourable outcomes, but we encountered a significant issue when transforming the model's output image from float to integers ranging from 0 to 255, resulting in a noticeable performance decline. It was also observed that utilizing ReLU in the output layer with an SSIM-based loss function resulted in overfitting to float values, causing train and validation SSIM to increase to 0.99. Nevertheless, upon transforming the output image to integers in the 0-255 range, the SSIM value declined sharply to approximately 0.7. Hence, we calculated the metrics in all our experiments after converting the generated image from float to int. The best results were observed using the Tanh activation function for the output layer.

The proposed model is trained with Adam optimizer \cite{adam} with a set learning rate of $10^{−4}$. The model is trained with different $\beta$ values ranging from 0 to 1, as indicated in Table 1, over the course of 500 epochs to explore their impact on the model's performance. Before training, the ground truth is normalized to a range of -1 to 1 to accommodate the model's Tanh activation function.
\begin{table}[h]
  \centering
  \begin{tabular}{ccc}
    \hline
    $\boldsymbol{\beta}$ & \textbf{Val. SSIM} & \textbf{Val. PSNR} \\
    \hline
    0 & 0.779 & 20.677 \\
    0.001 & 0.7417 & 20.198 \\
    0.01 & 0.7003 & 20.364 \\
    0.1 & 0.72 & 20.452 \\   
    1 & 0.7335 & 20.35 \\
    \hline

  \end{tabular}
\caption{Validation(Val.) Results for Various values of $\beta$ ( Hyperparameter for Contrastive Loss) when missing RGB Modality}
\label{tab:Table 3m}
\end{table}
During training, only horizontal and vertical flip augmentations are applied due to the given thermal image's low resolution, as more complex augmentations result in a significant loss of thermal information.

%We conducted experiments by varying the hyper-parameter $\beta$, using a single modality Low-Resolution Thermal image as input for the model. Our observations revealed that higher $\beta$ values lead to better regularization and improved results during training. However, this trend is inconsistent when a High-Resolution RGB image is missing. We noted that as the $\beta$ value increases, the model's ability to tolerate missing modality decreases, indicating a trade-off between regularization and missing modality robustness. Therefore, selecting the appropriate $\beta$ value to maximize the objective while carefully considering this trade-off is crucial.
We conducted experiments by varying the hyper-parameter $\beta$, using a single modality, i.e., either the Low-Resolution Thermal image or the high-resolution RGB image as input for the model. Our observations revealed that higher $\beta$ values lead to better regularization and improved results during training. However, this trend is inconsistent when a modality is missing. We noted that as the $\beta$ value increases, the model's ability to tolerate missing modality decreases, indicating a trade-off between regularization and missing modality robustness. Therefore, selecting the appropriate $\beta$ value to maximize the objective while carefully considering this trade-off is crucial.

Our primary evaluation metrics are Peak Signal-to-Noise Ratio (PSNR) and Structured Similarity Index Measure (SSIM). PSNR is preferred as it is simple to compute and is mathematically convenient in the context of optimization. However, it is sensitive to noise in the images and does not provide any information about the computational cost or efficiency of the super-resolution algorithm, which is needed for real-world applications and doesn't consider the structural similarity of the reconstructed image to the original image. SSIM is therefore preferred over PSNR\cite{5596999} as it is less sensitive to noise in the image. It provides a more accurate measure of the quality of the reconstructed image, along with a better measure of the perceptual quality of the image, and is more robust to changes in lighting and contrast. However, SSIM can be sensitive to relative scalings, translations, and rotations.
%\bibliography{egbib}
%\end{document}

\section{Results}
\begin{figure}[h]
  \centering
  \includegraphics[width=0.9\linewidth]{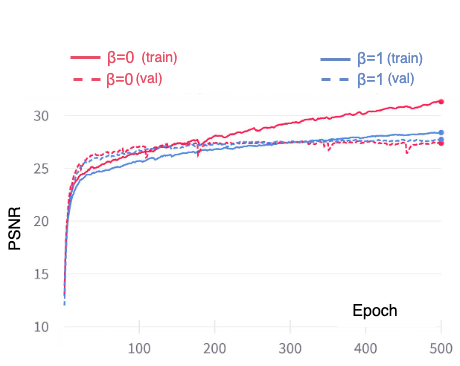}
  \caption{Comparing PSNR values for Train and Validation sets with different values of $\beta$}
  \label{psnr}
\end{figure}

\begin{figure}[h]
  \centering
  \includegraphics[width=0.9\linewidth]{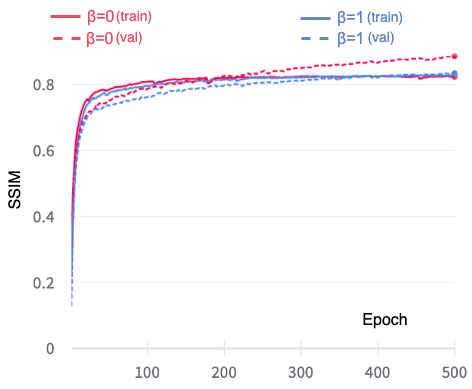}
  \caption{Comparing SSIM values for Train and Validation sets with different values of $\beta$}
  \label{ssim}
\end{figure}

Our experimentation shows that using $\beta$=1 results in the most favourable outcomes, as demonstrated in Table \ref{tab:Table 1}, with a validation SSIM score of 0.8322 and a PSNR score of 28.039, representing a noteworthy improvement compared to the outcomes obtained with $\beta$=0. 
However, in the absence of a modality, Table \ref{tab:Table 2} shows that the model's performance using only the low-resolution thermal resolution as input decreases when $\beta$=1 is utilized. The figures \ref{psnr} and \ref{ssim} display the regularization effect of adding contrastive loss, where we have logged the training and validation metrics for the model with $\beta$=0 and 1 for 500 epochs.
As observed in the results, our proposed model delivers promising outcomes even when the low-resolution thermal image is the only input, achieving an SSIM score of 0.75. With the addition of the high-resolution RGB image, the score increases up to 0.83. This indicates that the model is using the features from the RGB image as guidance and not overfitting on it. 
\begin{table}[h]
  \centering
  \begin{tabular}{ccc}
    \hline
    $\boldsymbol{\beta}$ & \textbf{Val. SSIM} & \textbf{Val. PSNR} \\
    \hline
    0 & 0.6107 & 11.4383 \\
    0.001 & 0.593 & 11.6168 \\
    0.01 & 0.5829 & 11.4913 \\
    0.1 & 0.5875 & 11.6262 \\   
    1 & 0.5714 & 11.3361 \\
    \hline
  \end{tabular}
\caption{Validation(Val.) Results for Various values of $\beta$ ( Hyperparameter for Contrastive Loss) when missing LR thermal Modality}
\label{tab:Table 2}
\end{table}

%------------------------------------------------------------------------
\section{Conclusion and Future Works}
\label{sec:formatting}
This paper presents a novel framework for data fusion and regularisation techniques for guided thermal image super-resolution using two encoders in a U-Net-based architecture. The proposed framework is our solution to the Thermal Image Super-Resolution Challenge - Track 2 at the 19th IEEE Workshop on Perception Beyond the Visible Spectrum. It is a simple, lightweight architecture that is computationally inexpensive compared to the other architectures pertaining to this domain. 

Our architecture gives the best result of 0.8322 SSIM and 28.039 PSNR score with the validation dataset when $\beta$ =1. The model possesses the ability to be robust in presence of missing data and maintain performance despite missing modalities, i.e., high-resolution RGB image or lower-resolution thermal image. To the best of our knowledge, we are the first to present a simple fusion model that is robust towards missing modalities, and these abilities make our architecture a computationally effective model for frequently occurring missing modalities in a real-world scenario.

In the future, we would like to explore domain-specific encoders for RGB and Thermal Images designed separately including transformer-based and other state-of-the-art segmentation architectures. We would also like to experiment with contrastive pretraining approaches along with other optimizations.

\section{Acknowledgments}
We would like to thank Mars Rover Manipal, an interdisciplinary student project team of MAHE, for providing the necessary resources for our research.

{\small
\bibliographystyle{ieee_fullname}
\bibliography{egbib}
}

\end{document}